\documentclass[manuscript,nonacm]{acmart} 

\AtBeginDocument{%
  }

\setcopyright{cc} 
\setcctype{by-nc-nd}
\copyrightyear{2026}
\acmYear{2026}

\begin{document}

\title{`AI': Ideologies of Computing}

\author{Andruid Kerne}
\email{andruid@ecologylab.net}
\affiliation{%
  \institution{Interface Ecology Lab, Department of Computer Science, University of Illinois Chicago}
  \country{USA}
}

\renewcommand{\shortauthors}{Kerne}

\begin{abstract}
We develop a conceptualization of ideology, in which a system of ideas represents social, economic, and political relationships.
We use ideology as a lens for understanding and critiquing intersecting social, economic, and political aspects of how `AI' technologies are being developed.
We observe ideological shifts.
We question that the present tangling of corporate and university objectives is beneficial to labor, particularly computer science students, and the general public.
\end{abstract}

\ccsdesc[500]{General and reference~Design}
\ccsdesc[500]{Human-centered computing}

\keywords{theory, human-computer interaction, epistemology, ideology, ai, future of work, labor, reflexivity, positionality}

\received{13 February 2026}
\received[revised]{24 March 2026}
\received[accepted]{11 March 2026}

\maketitle

\section{Introduction}
\label{introduction}
The development of `AI' is fueled by the positionalities of corporations, university computer science departments, students who graduate and enter the labor market, and the general public.
These positionalities take form through \emph{ideologies} that associate aspects of the social relations among these constituencies.
Clifford develops the reflexive formulation, that any ethnography is at least as much about its author as it is about its subjects \cite{clifford1986introduction}. So, too, what various actors say and do about `AI' is more about how their positionalities become represented in ideologies of business and science than it is about particular technologies.

We begin by developing a notion of what ideologies are and how they work. 
We start applying this conception to computer science and then to 
'AI'.
We develop an implication for design and a brief conclusion.


\section{Ideology (In the Context of Computing Institutions)}
What is ideology?
The OED provides, ``A systematic scheme of ideas, usually relating to politics, economics, or society and forming the basis of action or policy\ldots'' \cite{oed2025}. An ideology is a way of thinking that underlies social relations of economic production and various positionalities.

We develop a conceptualization of \emph{ideology} based on the scholarship of the Jamaican-British sociologist, Stuart Hall, who draws on Althusser \cite{althusser1969contradiction,althusser2024ideology}.
We contextualize this definition to address how computing functions in universities and corporations in the production of `AI' technologies.
According to Hall, ``Ideologies are the frameworks of thinking and calculation about the world---the `ideas' which people use to figure out how the social world works, what their place is in it and what they ought to do''~\cite{hall1985signification}[p.~99].
In other words, ideologies serve as the basis for the positionalities of actors, a grounding and source for their ideas about their roles in the economy and how those roles get intertwined.

Hall articulates how ideologies operate in relation to the social and the economic.
\begin{quote}
    What is the function of ideology? It is to reproduce the social relations of production. The social relations of production are necessary to the material existence of any social formation or any mode of production\ldots
    [Labor]  is 
    produced in the domain of the \emph{superstructures} [italics added]: in institutions like the family and church. It requires cultural institutions such as the media, trade unions, political parties, etc., which are not directly linked with production, as such, but which have the crucial function of "cultivating" labor of a certain moral and cultural kind---that which the modern capitalist mode of production requires\ldots
    [p. 98].
\end{quote}
By material, he means involving the artifacts that the economy produces. Where in his time artifacts were primarily analog and physical, in ours we focus on informational and computational artifacts. Among the institutions of the  \emph{superstructures}, along with family, church, and the media, we include the university, including its roles in education and in research.
Among forms of labor, we focus on computer science students and graduates, because they are vital in a market that is becoming transformed by `AI'.

A sense of \emph{difference} is key in the conceptualization of ideology that Hall derives by extending Althusser.
Difference is part of multivalent perspectives on and shifts among ideology forms.
\begin{quote}
    [Ideology] is a complex structure in which it is impossible to reduce one level of practice to another in some easy way\ldots
    The theorization of difference---the recognition that there are different social contradictions with different origins; that the contradictions which drive the historical process forward do not always appear in the same place, and will not always have the same historical effects [p. 91-92].
\end{quote}
By complexity, Hall means that there is not a direct causality between the social relations of production and ideology.
Formations can shift and evolve.
He develops the example of \emph{Black}, which is first used as a derogatory term in colonialism, but shifts to the site of \emph{Black Power}.
We see the roles of computer science students and graduates and the ideology of computer science shifting in a different direction.


\section{Shifting Ideologies of Computing}
The social relations of production have included that universities train students, that people who work for and at universities perform research, and that corporations hire students, graduates, and professors.
The ideology of education in universities is that knowledge and skills are transmitted to students in order to enable them to function as labor.
The ideology of research in universities is that we produce new knowledge that is beneficial to society, to corporations and the general public.
The ideology of computer science is that computers and networks empower people.
The operation of these ideologies is implicit and invisible.

We encounter the ideology of computing as empowerment in scholarship and media, as well as in practice.
Ted Nelson argued that through they could be used by corporations as tools of domination and boredom, that through hypertext, computers would become wonderful dream machines \cite{nelson1974computer}[p. 45].
Ridley Scott's Super Bowl ad introducing The Apple Macintosh situated it amidst footage of Orwell's \emph{1984}, and asserted that through it, ``You'll see why 1984 won't be like \emph{1984}'' \cite{scott1984apple}.
Retrospectively, we can see ironies here.
Ideology did not vanish, but instead, was hidden in plain sight, where personal computing devices meet networked platforms. As per Hall's theory of ideology,  different meanings emerge.
The role of popular personal computing devices---that is, the Macintosh and iPhone---as instruments of domination in society was foretold.
Hypertext social media platforms have been found to be addictive \cite{kang2026addictive}.

Because computing's role in society and the economy has become so central, computer science has become the most popular undergraduate major in many universities \cite{universities2023,nsc2024}.
Yet, we observe a transition in the role of computer science in the labor market, in which applying for jobs after college has shifted from abundance to relative scarcity.
According to Adarlo, as of December 2025, the unemployment rate for computer science graduates is greater than that for college grads as a whole \cite{adarlo2025dead}.
Getting internships, interviews, and offers has shifted from straightforward to difficult.

In popular parlance, CS grads are `cooked'.
This transformation of the social relations of production is problematic for students and graduates, but not for corporations.
Again, ideological interpretations shift.
University computer science is caught in the middle.
For example, in 2025-26, the UIC Computer Science department experienced its first ever year-to-year decrease in enrollment.
In the same year, after meany years of growth, construction, and planning, the department opened a beautiful new building.
Irony abounds.

\section{Ideologies of `AI'}
Interesting effects of ideologies are situated in the border zone \cite{kerneOpen2002,kerne2001collagemachineAModel} where computing, as mediated by interaction, meets the economy and the public through the rise of `AI'.
The rise of AI' has followed the rise of computer science in education, research, labor, and the economy.
This rise has been conducive not to questioning and critical thinking, which seem essential to research, but rather to cheerleading.

A  structural and ideological shift comes from the collapse of the ``border between research and products'' \cite{burrell2024introduction}. 
The prescription that Large Language Models (LLM)s must be as large as possible means that only large, intensively capitalized corporations are able to produce them. 
As a result, these uber LLMs are playing an outsized role in new research.
Burrell and Metcalf observe that the collapse and obsessive focus on scale problematizes, ``which research paths [the tight feedback loop between application and research] accelerates, which research paths it never considers, and who it benefits'' \cite{burrell2024introduction}.
When university researchers lose agency in defining the epistemologies of our research, such as in obeisance to the advancement without questioning of `AI', we become less able to serve society.

Gebru and Torres show how corporations' are obsessively focused on Artificial General Intelligence (AGI)---beyond LLMs-- and that this obsession is based on an ideology of eugenics and xenophobia \cite{gebru2024tescreal}.
They show how building well-defined systems instead of (obsessively) pursuing scale and AGI is required in order to be able to prioritize peoples' safety. We add that this is a requirement for serving humanity. 
A goal of the AGI project is to replace rather supplement labor. Who is this good for?

LLM technology functions as a boundary object \cite{star1989institutional}, which means different things to corporations, researchers, labor, and the public.
By calling it `AI', tech companies deliberately confuse the public about what the technology is, how it works, and what it is capable of.
OpenAI CEO Altman recently says, interacting with ChatGPT, ``\dots really feels like talking to a PhD-level expert'' \cite{heath2025gpt5}.
But LLMs like ChatGPT and Claude sometimes make obvious mistakes. They lack awareness of this. `Hallucination' is a euphemism.

This is why we put `AI' in quotes.
Altman's marketing flimflam combines with people's anxiety about not understanding and so falling behind.
The use of misleading signifiers compounds people's difficulties in comprehension.
In this ideology, the technology is exalted. People are pressured to catch up.

Somers develops an expose of the Claude LLM and its company, Anthropic \cite{somers2025anthropic}.
He admires the workers who left OpenAI, because it became too commercial, to join Anthropic.
He is excited that Anthropic employs philosophers and sociologists whose job is to characterize how Claude possesses a human-like character. But is this good or more sleight of hand?
The piece revels in Claude's idiosyncrasies.
What is missing is a fundamental epistemological critique of if Claude is in fact intelligent, and, further, if seeking to develop AGI through Claude is good for humanity.

\section{Implication for Design}
The corporate ideology of 'AI' and 'AGI' hurts students, graduates, and research.
An implication for design is that this ideology makes it nearly impossible, yet vitally important, to question the march of `AI' progress. 
Are LLMs `AI'? Is scaling LLMs and vying for AGI a good choice for society? 
Who will be served by devoting most of humanity's potential electricity production to technology that could reduce employment?
Asking these questions is not a route to getting the next grant or the most students.

Google Scholar finds that the most cited HCI research publishing venue is the flagship CHI Conference. This has been true for many years.
We observe that at this time, among the 20 most cited CHI publications of the past 5 years, 17 are about `AI', including 10 of the top 10 \cite{scholar2025chi}.
In 2019, addressing 2014-2019, we anecdotally recall that Google Scholar identified as 14 of the top 20 most cited CHI publications qualitative studies problematizing how people use technology, including harms as well as benefits.
Since the harms of technology to society have only increased over this period, this does not portend well.

Bender et al's stochastic parrots paper problematized the scale and energy consumption of LLMs \cite{bender2021dangers}.
The findings were obvious.
Yet, at the onset of its publication, Google fired Timnit Gebru, Margaret Mitchell, and other members of the Ethical `AI' team \cite{metz2020google,hao2020we} for declining to pull the paper from publication. 
This choice to try and suppress this critique of LLMs materially represents the ideology of Google.


\section{Conclusion}
Ideologies underpin the social relations of production.
Ideologies shift as part of shifts in the economic, technological, and social relationships among institutions, corporate and educational, labor, and the public. 
Calling LLMs `AI' is marketing sleight of hand that, over time, serves corporations better than computer science departments, especially computer science students.
Likewise the pursuit of unbounded AGI does not serve labor and the general public.
Contemporary phenomena involving the rise of `AI' and the fall of computer science can be understood in terms of ideological articulations and shifts as positionalities conflict.
So can the paucity of epistemological and existential questioning of how `AI' is affecting humanity.


\bibliographystyle{ACM-Reference-Format}
\bibliography{ideology,interfaceEcology,semioticsComputing}

\end{document}